\documentclass[conference]{IEEEtran}
\IEEEoverridecommandlockouts
\usepackage{cite}
\usepackage{lscape} 
\usepackage{longtable}
\usepackage{booktabs} 
\usepackage{multirow} 
\usepackage{amsmath,amssymb,amsfonts}
\usepackage{algorithmic}
\usepackage{graphicx}
\usepackage{textcomp}
\usepackage{subcaption}
\usepackage{ragged2e,array,booktabs}
\usepackage{siunitx}
\usepackage{array, ltablex, multirow}
\usepackage{tabularx}
\usepackage{colortbl}
\usepackage{hhline}
\usepackage{acronym}
\usepackage[latin9]{inputenc}
\usepackage[table]{xcolor}
\usepackage{collcell}
\usepackage{hhline}
\usepackage{pgf}
\usepackage{multirow}
\usepackage{soul}
\usepackage{tablefootnote}

\def\colorModel{hsb} 

\newcommand\ColCell[1]{
  \pgfmathparse{#1<50?1:0}  
    \ifnum\pgfmathresult=0\relax\color{white}\fi
  \pgfmathsetmacro\compA{0}      
  \pgfmathsetmacro\compB{#1/100} 
  \pgfmathsetmacro\compC{1}      
  \edef\x{\noexpand\centering\noexpand\cellcolor[\colorModel]{\compA,\compB,\compC}}\x #1
  } 
\newcolumntype{E}{>{\collectcell\ColCell}m{0.4cm}<{\endcollectcell}}  

\newcommand{\fakepar}[1]{\vspace{1mm}\noindent\textbf{#1.}}

\def\BibTeX{{\rm B\kern-.05em{\sc i\kern-.025em b}\kern-.08em
    T\kern-.1667em\lower.7ex\hbox{E}\kern-.125emX}}

\acrodef{adc}[ADC]{Analog-to-Digital Converter}
\acrodef{tmd}[TMD]{Transport Mode Detection}
\acrodef{keh}[KEH]{Kinetic Energy Harvesting}
\acrodef{seh}[SEH]{Solar Energy Harvesting}
\acrodef{teh}[TEH]{Thermal Energy Harvesting}
\acrodef{rfeh}[RFEH]{RF Energy Harvesting}
\acrodef{iot}[IoT]{Internet of Things}
\acrodef{rf}[RF]{Random Forest}
\acrodef{dt}[DT]{Decision Tree}
\acrodef{svm}[SVM]{Support Vector Machine}
\acrodef{knn}[KNN]{K-Nearest Neighbor}
\acrodef{nb}[NB]{Naive Bayes}
\acrodef{apr}[APR]{Acquisition Power Ratio}
\acrodef{led}[LED]{Light Emitting Diode}
\acrodef{cv}[CV]{Cross Validation}
\acrodef{rfe}[RFE]{Recursive Feature Elimination}
\acrodef{smote}[SMOTE]{Synthetic Minority Over-sampling Technique}
\acrodef{mpp}[MPP]{Maximum Power Point}
\acrodef{mosfet}[MOSFET]{Metal Oxide Semiconductor Field Effect Transistor}
\acrodef{pmu}[PMU]{Power Management Unit}
\acrodef{eno}[ENO]{Energy Neutral Operation}
\acrodef{hvac}[HVAC]{Heating, Ventilation, and Air Conditioning}
\acrodef{iv}[IV]{Current-Voltage Characteristic Curve}

\usepackage{refcount}
\begin{document}
\title{Towards Optimal Kinetic Energy Harvesting for the Batteryless IoT \\
}

\author{\IEEEauthorblockN{Muhammad Moid Sandhu\textsuperscript{1,2}, Kai Geissdoerfer\textsuperscript{3}, Sara Khalifa\textsuperscript{2,4}, Raja Jurdak\textsuperscript{5,2}, Marius Portmann\textsuperscript{1}, Brano Kusy\textsuperscript{2}}
\IEEEauthorblockA{\textsuperscript{1}School of Information Technology \& Electrical Engineering, The University of Queensland, Brisbane, Australia \\
\textsuperscript{2}Data61, Commonwealth Scientific \& Industrial Research Organization (CSIRO), Australia\\
\textsuperscript{3}Networked Embedded Systems Lab, TU Dresden, Germany\\
\textsuperscript{4}School of Computer Science \& Engineering, University of New South Wales, Sydney, Australia \\
\textsuperscript{5}School of Electrical Engineering \& Computer Science, Queensland University of Technology, Brisbane, Australia \\
{\small m.sandhu@uqconnect.edu.au, kai.geissdoerfer@tu-dresden.de, sara.khalifa@data61.csiro.au, r.jurdak@qut.edu.au,}\\ {\small marius@itee.uq.edu.au, brano.kusy@csiro.au}}
}

\maketitle

\begin{abstract}
Traditional \ac{iot} sensors rely on batteries that need to be replaced or recharged frequently which impedes their pervasive deployment. A promising alternative is to employ energy harvesters that convert the environmental energy into electrical energy. \ac{keh} converts the ambient motion/vibration energy into electrical energy to power the \acs{iot} sensor nodes. However, most previous works employ \ac{keh} without dynamically tracking the \textit{optimal} operating point of the transducer for maximum power output. In this paper, we systematically analyze the relation between the operating point of the transducer and the corresponding energy yield. To this end, we explore the voltage-current characteristics of the \ac{keh} transducer to find its \ac{mpp}. We show how this operating point can be approximated in a practical energy harvesting circuit. We design two hardware circuit prototypes to evaluate the performance of the proposed mechanism and analyze the harvested energy using a precise load shaker under a wide set of controlled conditions typically found in human-centric applications. We analyze the dynamic current-voltage characteristics and specify the relation between the \ac{mpp} sampling rate and harvesting efficiency which outlines the need for dynamic \ac{mpp} tracking. The results show that the proposed energy harvesting mechanism outperforms the conventional method in terms of generated power and offers at least one order of magnitude higher power than the latter.
\end{abstract}
\vspace{-0.4cm}
\section{Introduction}

The \ac{iot} is identified as one of the major technologies impacting various applications including healthcare, agriculture, transportation, smart homes and more, by enabling many sensors to collect and exchange data for smarter and effective decisions~\cite{lin2017survey}.
However, \ac{iot} sensors rely mostly on batteries for their operation. Batteries are bulky, expensive, hazardous and have limited charge/discharge cycles which impedes the pervasive deployment of \ac{iot}~\cite{hester2017future}.
As a way of tackling these challenges, energy harvesting from ambient sources such as solar, kinetic, thermal and RF waves has been used to convert the environmental energy into electrical energy to power batteryless \ac{iot} sensors without human intervention~\cite{lin2017survey}.
With recent technological advancements, energy harvesting can enable perpetual operation of \ac{iot} sensor nodes which is the key for realising a truly pervasive, batteryless \ac{iot}.

\begin{figure}[t!]
\centering
\includegraphics[width=8cm, height=4cm]{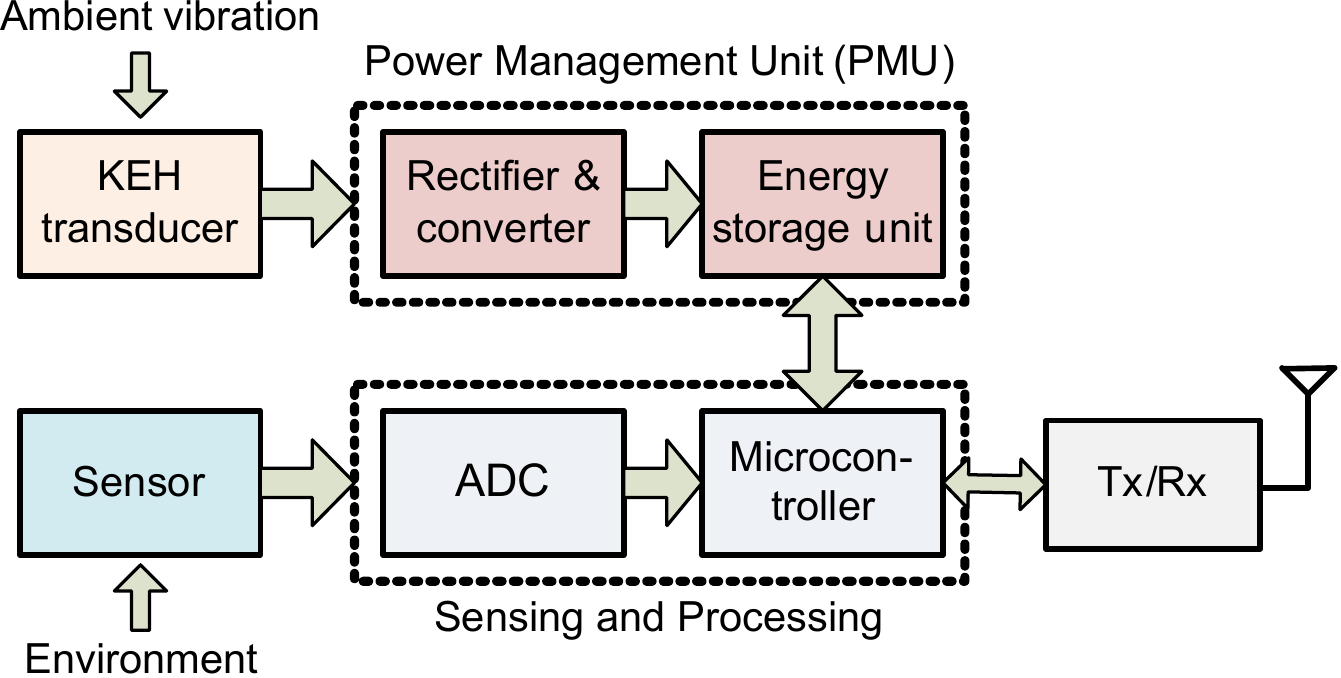}
\caption{Energy harvesting mechanism to power the \ac{iot} nodes}
\label{fig:intro_iots}
\vspace{-0.65cm}
\end{figure}

\acl{keh} is an attractive option to realise a truly pervasive batteryless \ac{iot} by converting the ambient vibration, motion and mechanical energy into electrical energy.
A \ac{pmu} is used to manage the harvested energy~\cite{estrada2018multiple} to power a load (or a node) as depicted in Fig.~\ref{fig:intro_iots}.
The harvested energy is stored in an energy storage unit (i.e., a capacitor/battery) which is used to power a sensor node.
The harvested energy is consumed by various hardware components within the sensor node including the sensor module, \ac{adc}, microcontroller and transceiver to sample the physical attribute as well as its transmission to the destination/server.
Depending upon the application, the harvested energy can be large enough to power the sensors without the need of any external depletable energy source leading to \ac{eno}. 

Piezoelectric transducers are often used to power the sensors in human-centric \ac{iot} applications including the human activity~\cite{kuang2017energy} and transport modes.
In order to harvest maximum energy from the transducer, it must operate at its \ac{mpp}. 
Although, \ac{mpp} tracking has been used in \ac{keh}~\cite{kuang2017energy} to increase harvested energy, it has not been systematically compared with the conventional design in terms of harvested energy.
In addition, the dynamic tracking of \ac{mpp} of the \ac{keh} transducer is still unexplored.
Furthermore, there is no detailed study to compare the performance of the energy harvesting circuits with and without \ac{mpp} tracking of the transducer under a wide set of conditions typically found in human-centric applications.

In this paper, we explore the operation of a \ac{keh} transducer at the \ac{mpp} using a DC-DC boost converter at different frequency and amplitude levels for human-centric \ac{iot} applications.
A DC-DC boost converter consumes some amount of energy during its operation which may impact the overall harvested energy compared to a converter-less design.
In this document, we refer to DC-DC boost converter as DC-DC converter and converter interchangeably.
In order to do a fair comparison, we perform the experiments in a controlled environment using a stable load shaker whose vibration frequency and amplitude can be adjusted manually.
We study the relation between harvesting efficiency and \ac{mpp} sampling rate which outlines the need for dynamic \ac{mpp} tracking.
The results show that exploiting the \ac{mpp} of the transducer, the proposed mechanism harvests, on average, at least one order of magnitude higher power, than the conventional converter-less design, depending on the vibration amplitude.
\begin{figure}[t!]
\centering
\includegraphics[width=8cm, height=3cm]{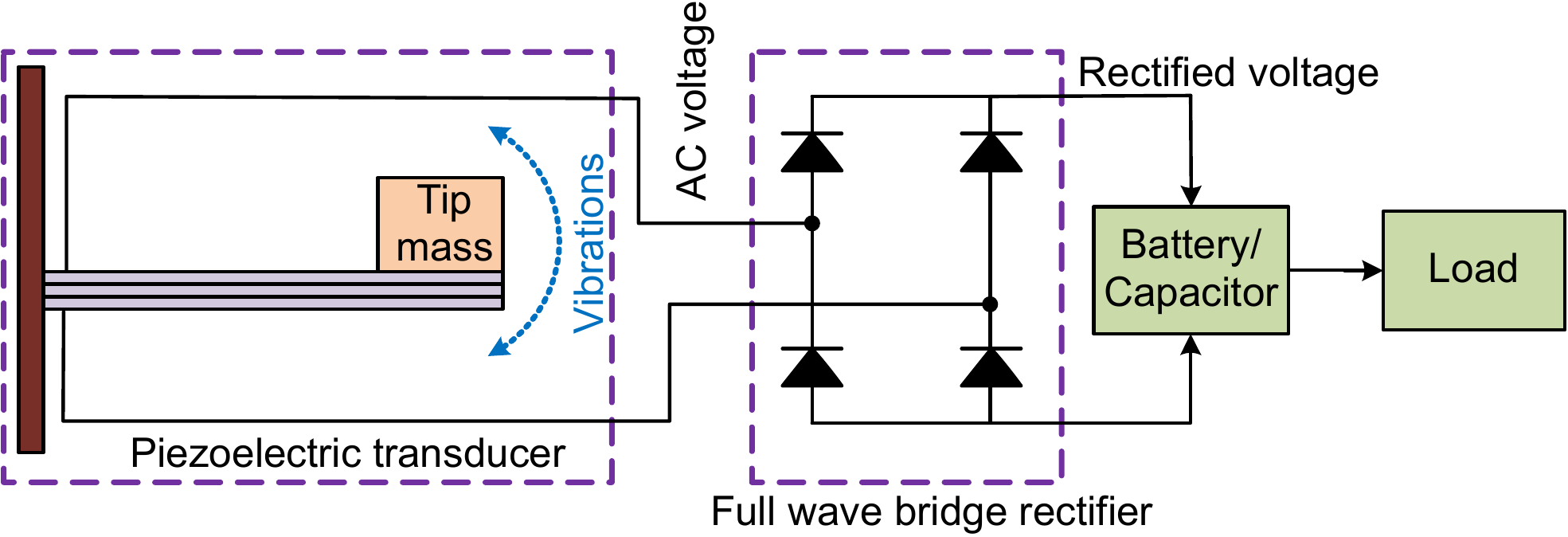}
\caption{General piezoelectric energy harvesting circuit}
\label{piezoelectric_transducer}
\end{figure}

The main contributions of the this paper are given as:
\begin{itemize}
    \item We analyze the dynamic electrical characteristics of a \ac{keh} transducer and explore its \ac{mpp}.
    \item We design two hardware circuit prototypes; converter-less and converter-based, to explore the impact of \ac{mpp} tracking using a DC-DC converter.
    \item We perform an experiment in a controlled environment using a stable load shaker and find that the proposed converter-based circuit harvests, on average, one order of magnitude higher power than the conventional converter-less design due to its operation at \ac{mpp}.
\end{itemize}

The remainder of the paper is organized as follows. Section~\ref{Literature_Review} discusses the previous energy harvesting based \ac{iot} devices and Section~\ref{Energy_Harvesting_Circuits} presents the proposed energy harvesting design. The prototyping and experimental setup is described in Section~\ref{prototyping}. Section~\ref{results} explores the \ac{mpp} of the \ac{keh} transducer and provides the results. Finally, Section~\ref{Conclusion_and_Future_Work} concludes the paper and gives the future direction.

\section{Related Work}
\label{Literature_Review}
In batteryless \ac{iot} nodes, energy harvesters are employed to provide unceasing energy for their operation.
Zhao et al.~\cite{zhao2014shoe} design a \ac{keh} based shoe to harvest energy from the human motion. Their prototype harvests \SI{1}{\milli\watt} of power while walking at a frequency of about \SI{1}{\hertz}. Their work shows the feasibility of using energy harvesters to power the batteryless wearable devices. Authors in~\cite{magno2016kinetic} design a \ac{keh} based wearable device to be placed on the wrist to harvest energy from human motion. As human arms offer lower vibrations compared to legs, they harvest \SI{280}{\micro\joule} of energy during small human movements. Ryokai et al.~\cite{ryokai2014energybugs} design a \ac{keh} based wearable device for children to harvest energy from their movements. The harvested energy from the wrist based wearable devices is used to turn on the \ac{led} to inspire the children to learn about energy and its transformation from one form to another. Olivo et al.~\cite{olivo2010kinetic} propose a fast start-up mechanism for wearable sensors using \ac{keh}. They employ a boosting circuit to step-up the harvesting voltage to power the wearable nodes. Authors in~\cite{xie2014human} employ \ac{keh} to harvest energy from human movements to power the wearable devices. They design backpack-based and insole-like harvesters which generate large enough energy to power the on-body nodes. Xiang et al.~\cite{xiang2013powering} use \ac{keh} transducer to harvest energy from the air flow of indoor \ac{hvac} system. The harvested energy is used to transmit information about the air flow speed to the server. 
Kuang et al.~\cite{kuang2017energy} employ \ac{keh} to harvest energy from the knee-joint energy harvester during human motion. They design \ac{keh} hardware which operates the transducer at its \ac{mpp} to harvest higher energy to power the body sensor.

\begin{figure}[t!]
\centering
\includegraphics[width=9cm, height=3.5cm]{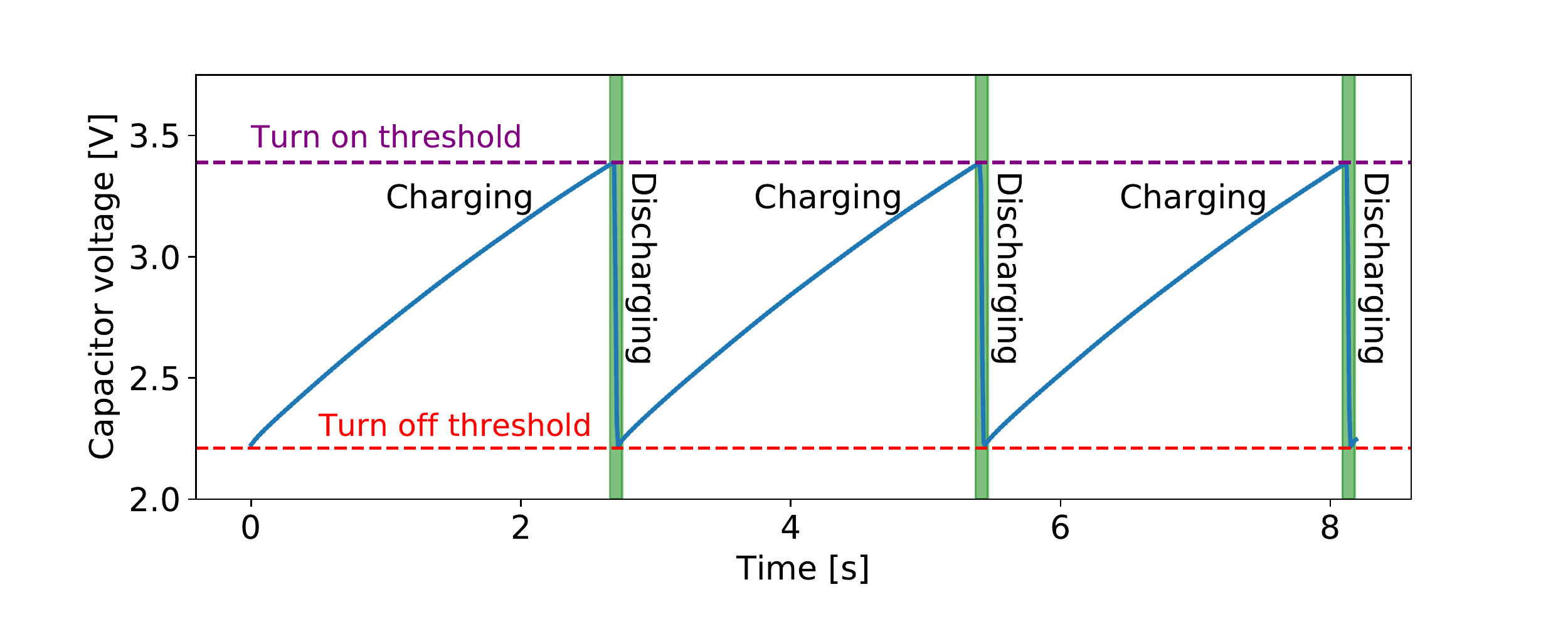}
\caption{Capacitor voltage in intermittently powered sensors}
\label{fig:transiently_powered_sensors}
\end{figure}

Some of the previous works~\cite{xiang2013powering, magno2016kinetic, olivo2010kinetic} employ DC-DC boost converters in the KEH circuits to harvest higher energy for \ac{iot} nodes. However, they do not explore the \ac{mpp} of the \ac{keh} transducer which is challenging due to the rapid changes in its output power.
The authors in~\cite{kuang2017energy} do not employ dynamic tracking of \ac{mpp} under a wide set of conditions typically found in human-centric applications.
Furthermore, to the best of our knowledge, there is no detailed study that compares the performance of employing the \ac{mpp} of the \ac{keh} transducer with the conventional energy harvesting mechanisms for miniaturized \ac{iot} sensor nodes.
This papers presents the operation of a \ac{keh} transducer at \ac{mpp} to harvest higher energy and compares its performance with the conventional converter-less energy harvesting circuit under a wide set of conditions typically found in human-centric applications.


\section{Kinetic Energy Harvesting}
\label{Energy_Harvesting_Circuits}
There are three mechanisms for converting the kinetic energy to electrical energy: piezoelectric, electromagnetic and electrostatic~\cite{khalifa2017harke}.
Piezoelectric is the most favourable choice for wearable applications due to its simplicity and compatibility with micro-electro-mechanical systems. We discuss the functionality of the piezoelectric transducer and the proposed energy harvesting circuits in the following subsections in detail.
\begin{figure}[t!]
\centering
\includegraphics[width=8.5cm, height=4cm]{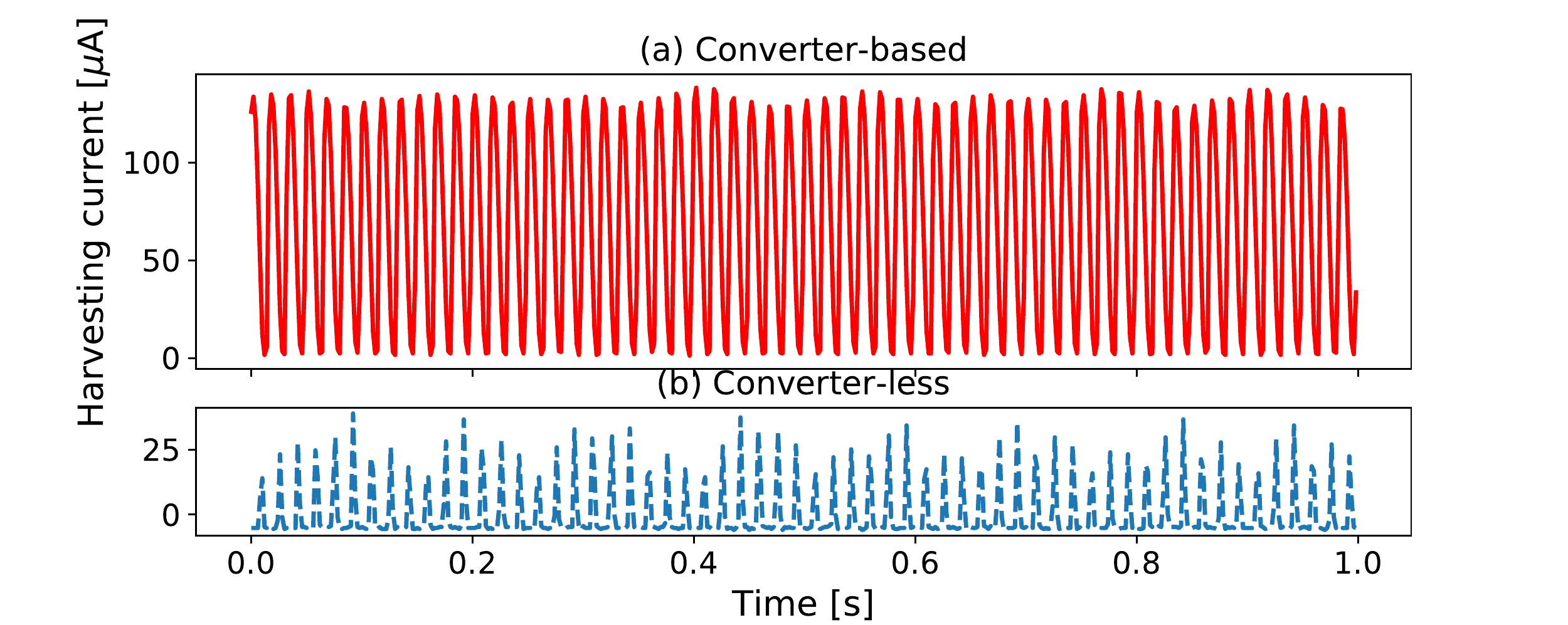}
\caption{Harvesting current in converter-less and converter-based circuits at the resonance frequency of the transducer}
\label{fig_interference_effect}
\end{figure}
\subsection{Piezoelectric transducer}
\label{sec:piezoelectric_transducer}
A piezoelectric transducer is used to convert the ambient vibration and kinetic energy into electrical energy as shown in Fig.~\ref{piezoelectric_transducer}.
Under mechanical stress, it generates an electric field with alternating polarity~\cite{khalifa2017harke}. The harvesting AC voltage is rectified using a bridge rectifier and is used to charge a capacitor. Then, the energy stored in the capacitor is used to power a load as shown in Fig.~\ref{piezoelectric_transducer}.
The harvested energy from the piezoelectric transducer depends upon many factors including the resonance frequency, amplitude and frequency of vibration.
The resonance frequency is an intrinsic property of the piezoelectric transducer, depending on its geometry, manufacturing parameters and material properties which can not be changed later.
Using a tip mass, the resonance frequency of the harvester can be tuned according to specific application requirements.
However, it requires human intervention each time while there is a change in vibration characteristic.
The amount of energy harvested from a piezoelectric transducer depends upon the dynamic vibration characteristics which in turn depends upon the application in which the transducer is being deployed.
These dynamics also change the optimum operating point of the transducer in terms of mechanical tuning as well as electrical loading.
Therefore, there is a need to devise a method that harvest maximum energy under given vibration characteristics without human intervention (i.e., without replacing piezoelectric transducers and/or tuning the resonance frequency). Accordingly, we explore the impact of these uncontrollable parameters (vibration amplitude and frequency) on the harvested energy using different energy harvesting hardware designs employing the \ac{mpp} of the transducer, as described in the following subsections and find the best design under a given nature of vibration characteristics. 

\subsection{Energy harvesting circuits using \ac{keh} transducer}
In batteryless \ac{iot} devices, the load is turned on when the capacitor voltage reaches to a turn-on threshold $V_{on}$ and is turned off when the capacitor voltage drops to a turn-off threshold $V_{off}$ as shown in Fig.~\ref{fig:transiently_powered_sensors}. The stored energy in the capacitor $E_{cap}$ is calculated as:
\begin{equation}
    E_{cap} = \frac{1}{2}CV^{2}
    \label{eq:cap_energy}
\end{equation}
where, $C$ is the capacitance and $V$ is the capacitor voltage. Fig.~\ref{fig:transiently_powered_sensors} shows the typical intermittent execution pattern of a batteryless device. The energy consumed by the load $E_{load}$ is given as:
\begin{equation}
    E_{load} = \frac{1}{2}C(V_{on}^{2}-V_{off}^{2})
    \label{eq:load_energy}
\end{equation}
As Fig.~\ref{fig:transiently_powered_sensors} depicts, the frequency of executing the load depends on the rate of charging of the capacitor.
There are two main design options for charging the capacitor which are described in detail below.
\begin{figure}[t!]
\centering
\includegraphics[width=9cm, height=4cm]{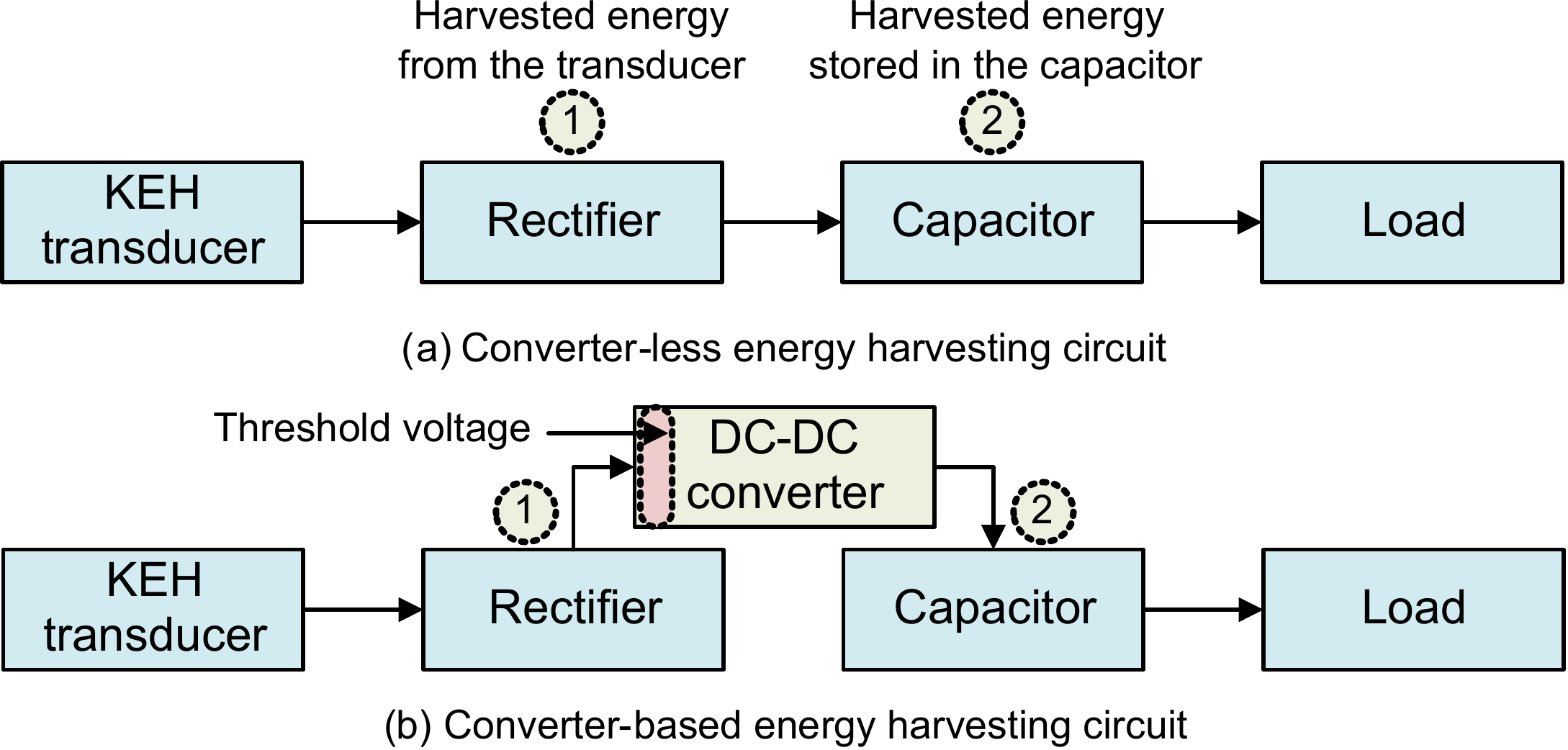}
\caption{Illustration of \ac{keh} hardware prototypes (a) Converter-less and (b) Converter-based designs}
\label{sensing_harware}
\end{figure}

\fakepar{Converter-less design}
This is a conventional circuit for harvesting energy and powering a load. In a converter-less design, the capacitor is placed in parallel to the rectifier.
Therefore, the voltage across the transducer is given as:

\begin{equation}
    v_{AC} = v_{cap}+2\cdot v_d
    \label{eq:envel_formation}
\end{equation}
where, $v_{cap}$ is the capacitor voltage and $v_d$ is the voltage drop across the corresponding diode in the full wave rectifier.
If the open circuit voltage is higher than the voltage on the capacitor, $v_d$ is approximately constant (one diode drop, typically \SI{350}{\milli\volt}\footnote{Using schottky diodes}) and thus, the voltage across the transducer equals the capacitor voltage plus two diode drops.
This may result in low energy yield.
For example, if the capacitor voltage is \SI{3}{\volt} and the input vibration is low, then the open circuit voltage of the transducer may be less than \SI{3}{\volt}$+2v_d$.
In this case, no current can flow and thus energy that could have potentially been harvested is wasted.
The dependence between capacitor voltage and transducer voltage has important implications for the harvested energy.


\fakepar{Converter-based design}
In a converter-based design, a DC-DC boost-converter~\cite{xiang2013powering} is placed between the rectifier and the capacitor.
This allows to optimize the operating point (i.e., harvesting voltage) of the transducer independent of the voltage on the capacitor.
For example, the converter can be configured to regulate the transducer voltage at its input to \SI{300}{\milli\volt} by dynamically controlling the current flow from the transducer.
This allows to extract energy from the transducer for charging the capacitor even under very low motion or vibrations. It can also be used to fix the transducer voltage to its \ac{mpp}, extracting higher energy from the transducer.
The decoupling of the transducer from the capacitor and load has a significant impact on the current flow from the transducer. Fig.~\ref{fig_interference_effect} shows the harvesting current from converter-less and converter based designs. In the converter-based design, the capacitor is decoupled from the transducer using DC-DC converter with fixed input voltage threshold according to \ac{mpp} of the transducer which allows four times higher current flow than the converter-less design.


\begin{figure}[t!]
\centering
\includegraphics[width=8cm, height=4cm]{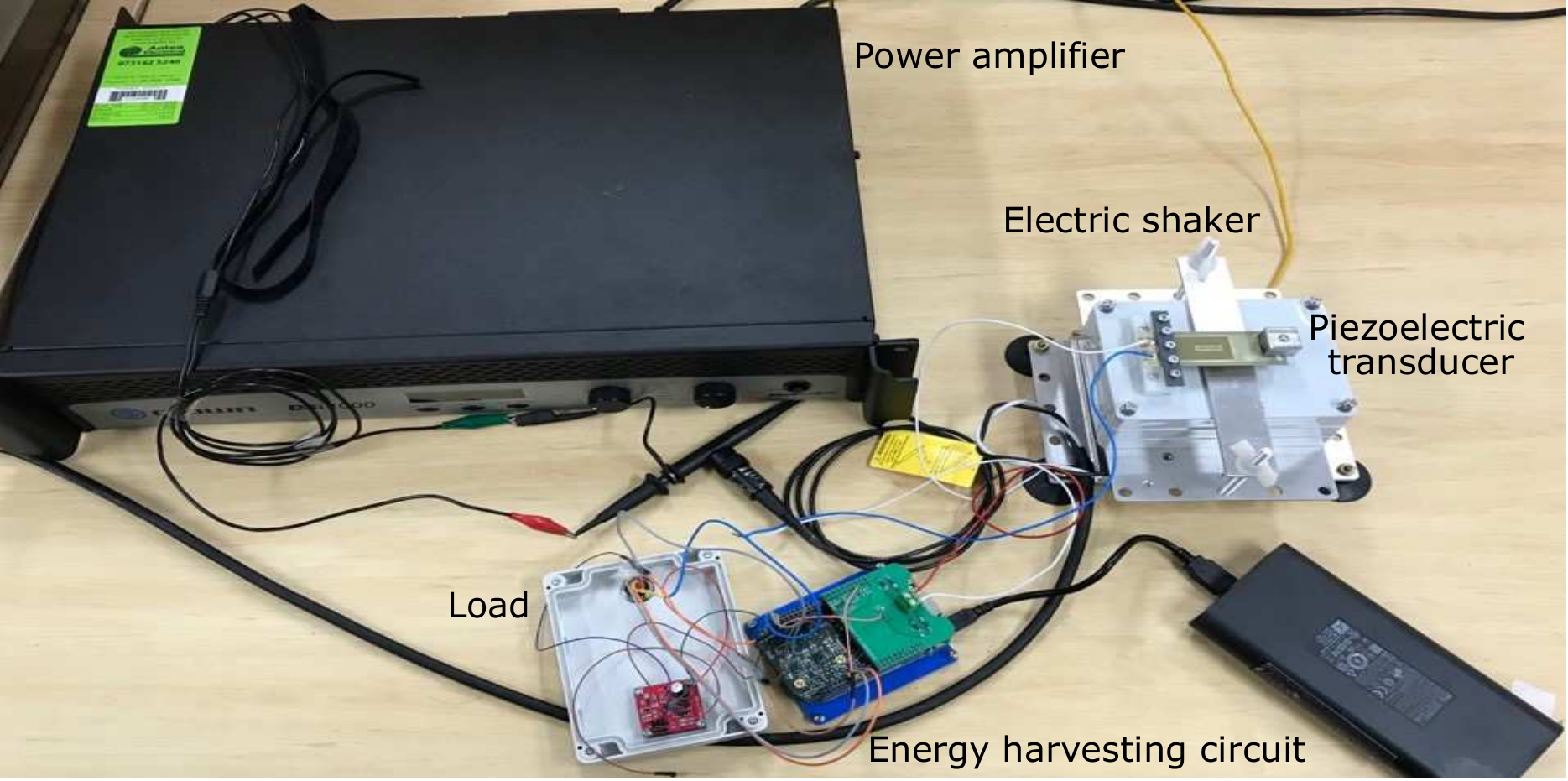}
\caption{Experimental setup for measuring the harvested power using KEH converter-less and converter-based hardware designs with a stable load shaker}
\label{fig:experiment_setup_smart_shaker}
\end{figure}

\section{Prototyping and Experimental Setup}
\label{prototyping}
For our study, we design two \ac{keh} hardware prototypes for collecting data from two types of energy harvesting circuits: converter-less (Fig.~\ref{sensing_harware}a) and converter-based (Fig.~\ref{sensing_harware}b).
We use an S230-J1FR-1808XB two-layered piezoelectric bending transducer from MID\'E technology\footnote{https://www.mide.com/} with a block tip mass of \SI{24.62}{g}\,$\pm$\,0.5\% which has a resonance frequency of \SI{25}{\hertz}.
All signals are sampled at a frequency of \SI{100}{\hertz} using a 12-bit \ac{adc}.
Both designs in Fig.~\ref{sensing_harware} use a \SI{220}{\micro\farad} capacitor to temporarily store the harvested energy and an intermittently powered load\footnote{In this experiment, we set $V_{on}$ = \SI{3.38}{\volt} and $V_{off}$ = \SI{2.18}{\volt}} consisting of two \ac{led}s, mimicking the behaviour of a batteryless device. The converter-based design (in Fig.~\ref{sensing_harware}b) uses an ultra low power TI BQ25504 DC-DC boost converter to step up the transducer voltage. The input threshold voltage of the converter is linearly swept from \SI{500}{\milli\volt} to \SI{2900}{\milli\volt} in \SI{100}{\milli\volt} steps.
The prototypes are designed as dataloggers with a focus on enabling accurate measurements of energy rather than on optimizing the harvesting efficiency.

We use a 1-lb load shaker\footnote{\label{note1}https://controlledvibration.com/product-item/1lb-load-shaker/} along with a power amplifier\footnotemark[\getrefnumber{note1}], DSi 1000, to excite the transducer at varying amplitude and frequency of vibration.
An arbitrary waveform generator\footnote{https://www.rigolna.com/products/waveform-generators/dg4000/}, RIGOL DG4162, is used to provide a single tone sinusoidal signal to the power amplifier which drives the load shaker.
The experimental setup for measuring the harvested power in converter-less and converter-based designs is shown in Fig.~\ref{fig:experiment_setup_smart_shaker}. In this experiment, we fixed the signal from the waveform generator to a sine wave with varying amplitude levels (referred as vibration amplitudes in the rest of the paper) from \SI{200}{\milli\volt} to \SI{1000}{\milli\volt} peak-to-peak and frequency from \SI{10}{\hertz} to \SI{50}{\hertz}. These values of vibration amplitude and frequency are selected by considering the real vibration characteristics from various transport modes~\cite{hemminki2013accelerometer}. A piezoelectric transducer is mounted on the load shaker, and the converter-less and converter-based energy harvesting circuits are employed consecutively to sample the generated voltage and current signals.
\begin{figure}[t!]
\centering
\includegraphics[width=9cm, height=3.5cm]{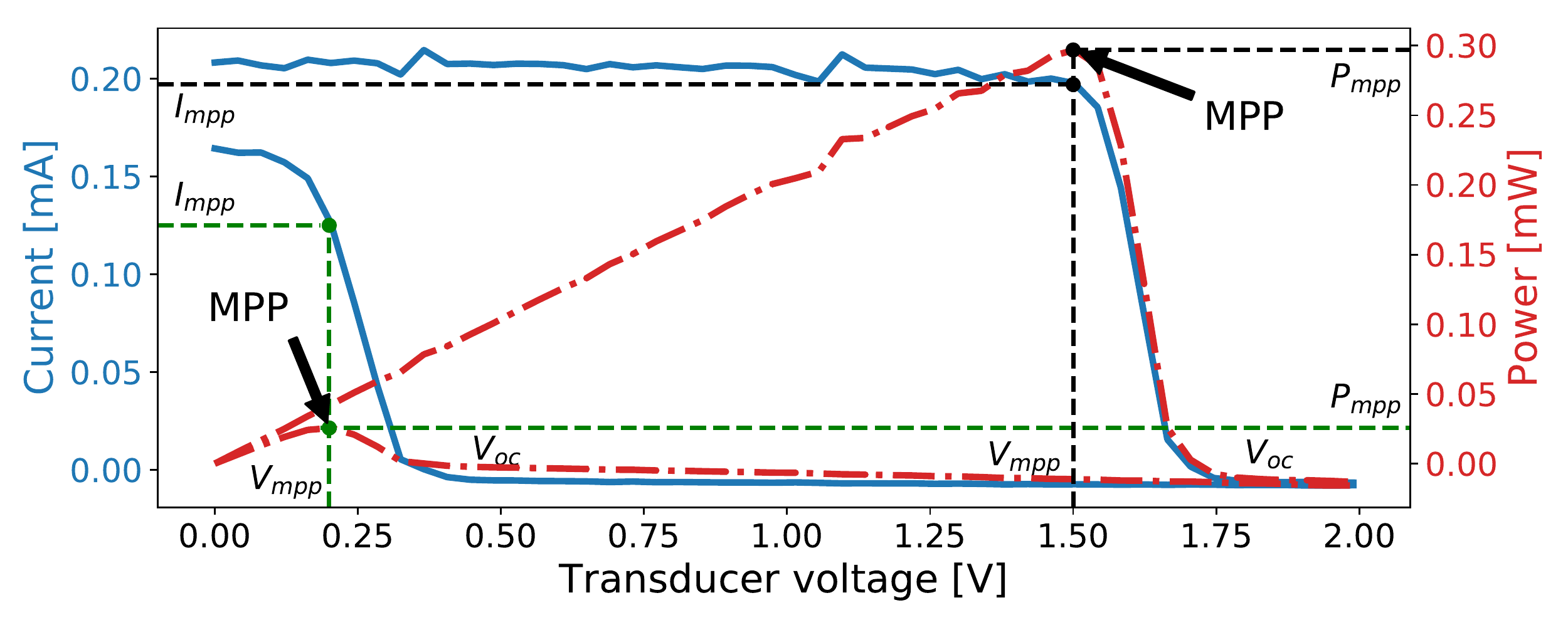}
\caption{The \ac{iv}, corresponding power and \ac{mpp} of the piezoelectric transducer operating at the resonance frequency}
\label{fig:piezo_iV_curve_mppt}
\end{figure}

\begin{figure}[t!]
\centering
\includegraphics[width=\linewidth]{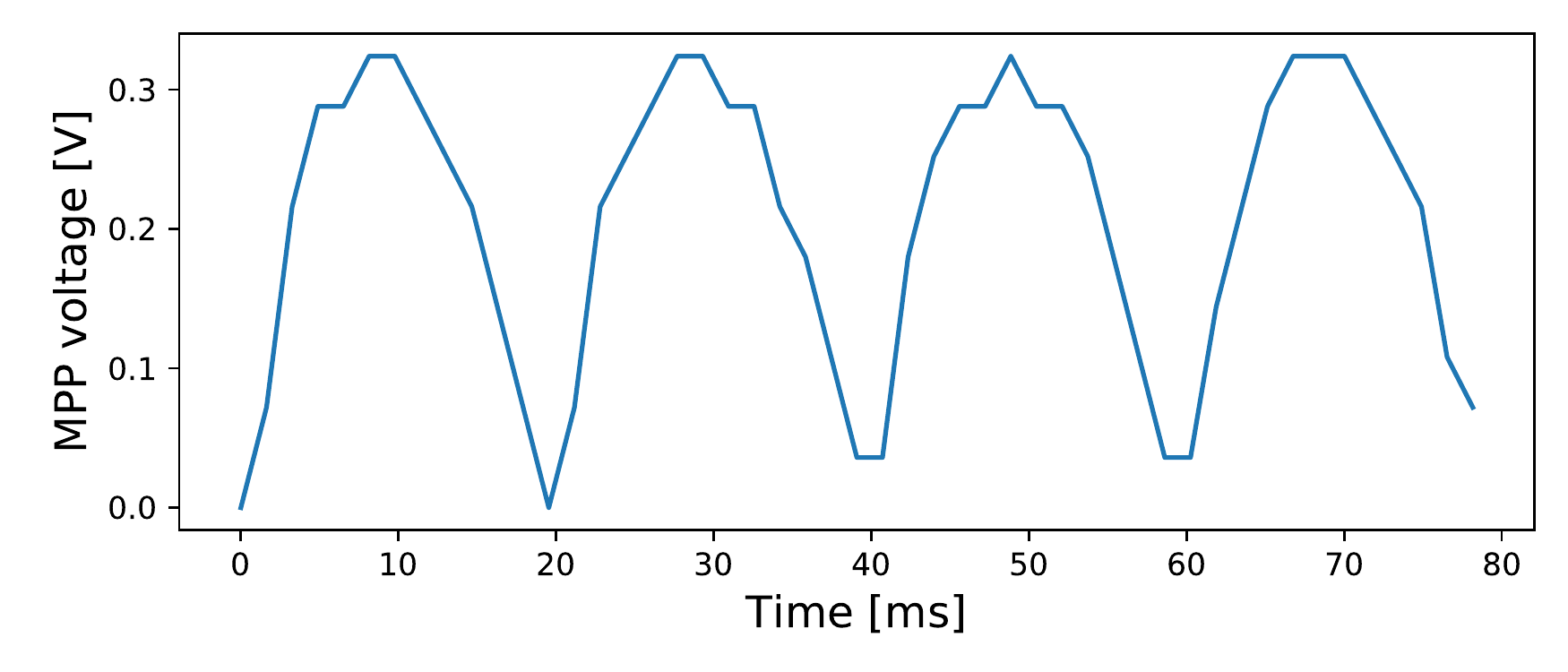}
\caption{The \ac{mpp} voltage changes with the oscillation of the transducer.}
\label{fig:mpp_voltage_with_time}
\end{figure}

\section{Results and Discussion}
\label{results}
In this section, we present the harvested power from the converter-less and converter-based designs, and study the benefit of operating the \ac{keh} transducer at its \ac{mpp}.
\subsection{\ac{mpp} of the \ac{keh} transducer}
\label{subsec:mpp_transducer}
In the converter-based energy harvesting circuit, we can choose the input voltage threshold of the DC-DC converter in order to operate the transducer at its \ac{mpp}.
As shown in Fig.~\ref{fig:piezo_iV_curve_mppt}, the \ac{mpp} of the piezoelectric transducer is defined by its current-voltage characteristics.
We explore the \ac{iv} characteristics of the piezoelectric transducer to find its \ac{mpp} using an uncontrolled bridge rectifier and a KEITHLEY 2602B source meter\footnote{https://www.tek.com/keithley-source-measure-units/smu-2600b-series-sourcemeter}. We rapidly sweep through the range of voltage and record the corresponding current at a sampling frequency of \SI{15.4}{\kilo\hertz}. 
In order to analyze the \ac{mpp} of the piezoelectric transducer, we plot two \ac{iv} curves in Fig.~\ref{fig:piezo_iV_curve_mppt}. It shows that the current from the transducer decreases with the increase in the output voltage, reaching to zero at the open circuit voltage $v_{oc}$. Fig.~\ref{fig:piezo_iV_curve_mppt} also plots the output power over the transducer voltage. It depicts that the \ac{mpp} lies near 50-80\%  of the open circuit voltage.
\begin{figure}[t!]
\centering
\includegraphics[width=\linewidth]{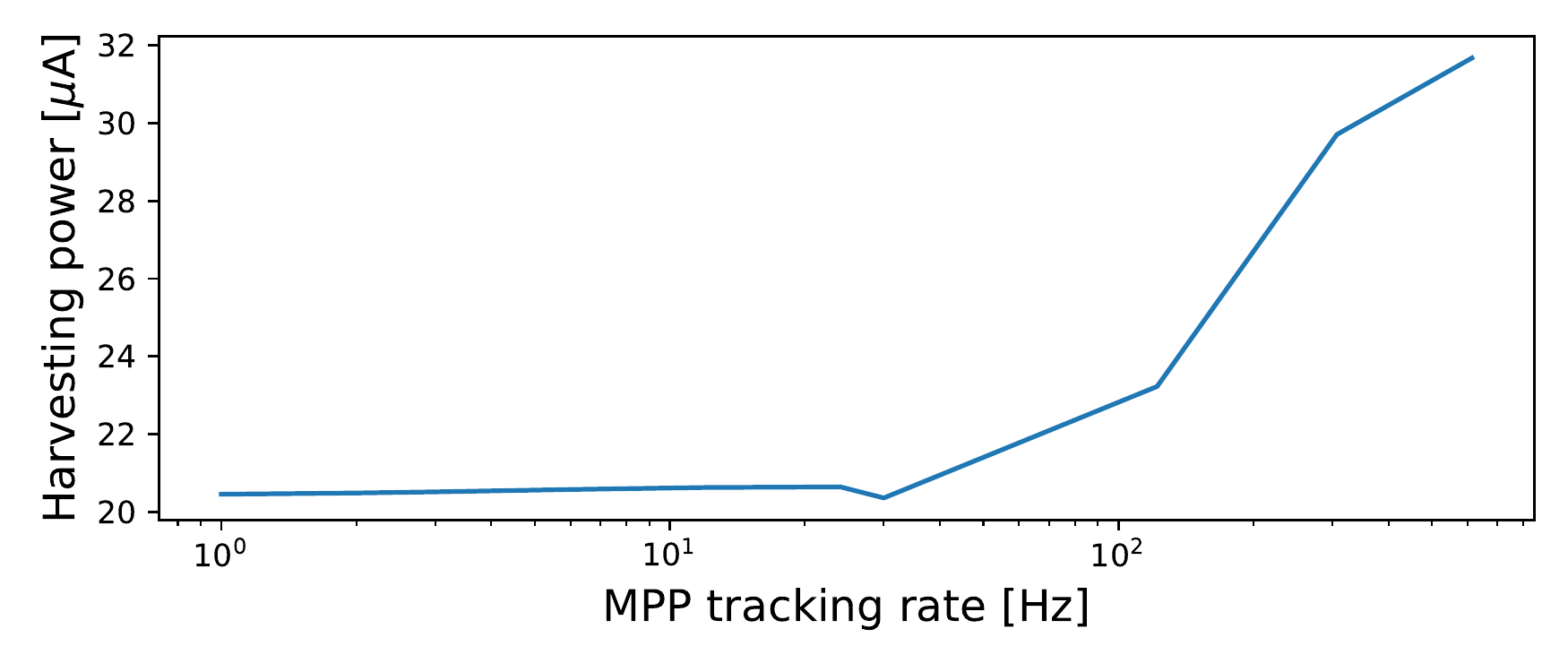}
\caption{The average transducer power depends on the \ac{mpp} sampling frequency}
\label{fig:average_power_frequency}
\vspace{-0.5cm}
\end{figure}
It also shows the corresponding current $I_{mpp}$ and power $P_{mpp}$ at \ac{mpp}. From the figure, it is clear that with the change in \ac{iv} curve, the \ac{mpp} also changes according to the open circuit voltage. Fig.~\ref{fig:mpp_voltage_with_time} shows the dynamic pattern of \ac{mpp} voltage with time.
The \ac{mpp} varies with the oscillations of the transducer due to the change in the output voltage.
We analyse the effect of slow \ac{mpp} tracking by calculating the power when undersampling the \ac{mpp},
i.e., we select the harvesting voltage at the \ac{mpp} and then keep harvesting at that voltage for $1/f_T$ seconds, where $f_T$ is the tracking frequency.
Fig.~\ref{fig:average_power_frequency} shows that harvested power significantly increases with increased \ac{mpp} tracking speed.
This means, that in order to extract maximum power from the transducer, we would need to dynamically track its \ac{mpp}.
Practical \ac{mpp} tracking circuits usually disconnect the harvester and sample the open circuit voltage in order to estimate the \ac{mpp}.
This incurs significant losses while the harvester is disconnected.
These losses linearly increase with increasing \ac{mpp} sampling rate.
Furthermore, off-the-shelf DC-DC converters, like the one used in our prototype offer only a fixed \ac{mpp} sampling interval of \SI{16}{\second}, which is far too low to track the \ac{mpp} of a \ac{keh} transducer.
Therefore, for this study, we only consider static \ac{mpp} tracking, where the \ac{mpp} is estimated once for the given vibration characteristics and then kept constant.

\begin{figure}[t!]
\centering
\includegraphics[width=9cm, height=5.5cm]{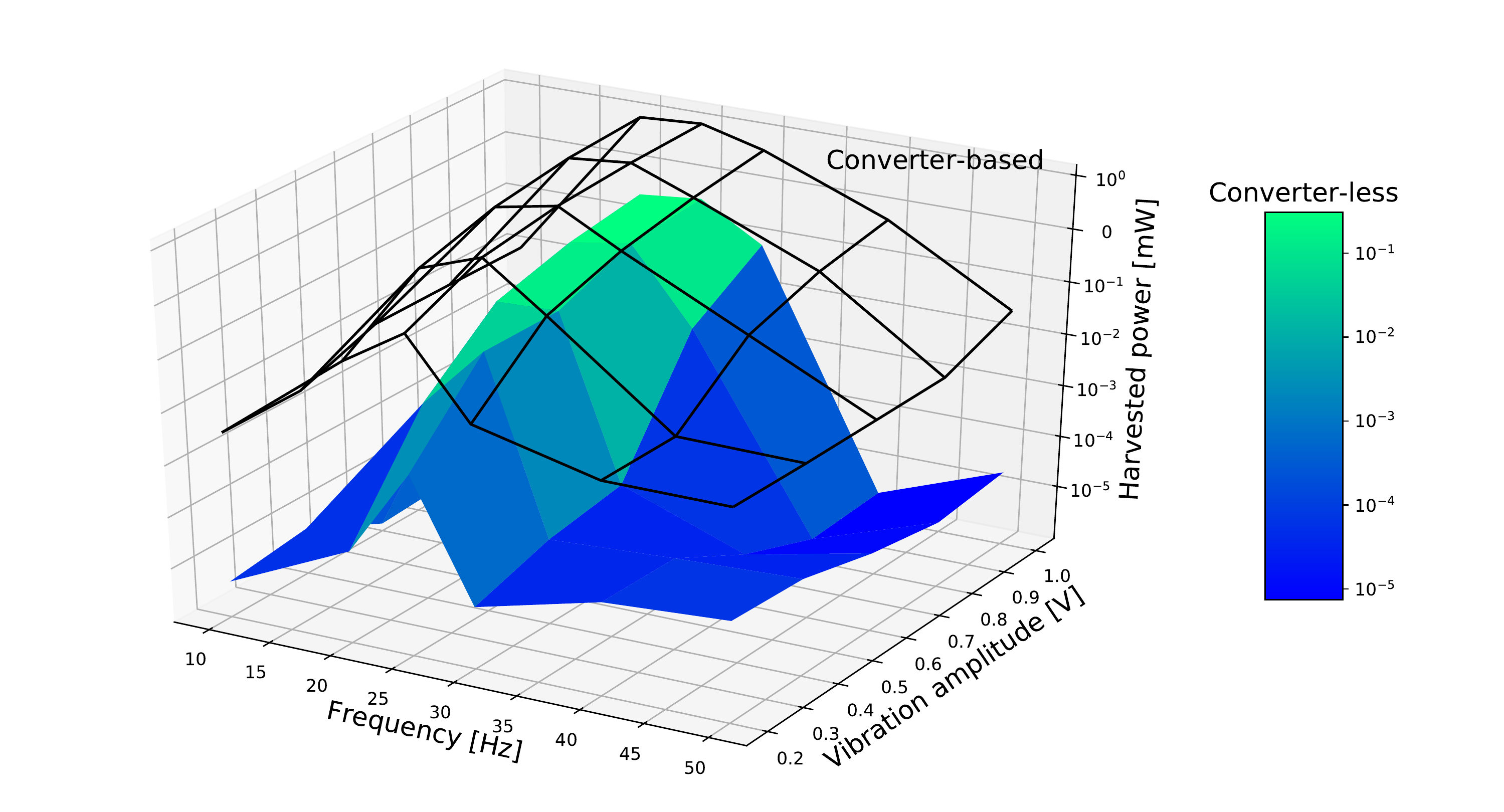}
\caption{The harvested power in converter-less and converter-based energy harvesting circuits}
\label{fig:CL_CB_comparing_shaker}
\end{figure}

\begin{figure}[t!]
\centering
\includegraphics[width=9cm, height=5.5cm]{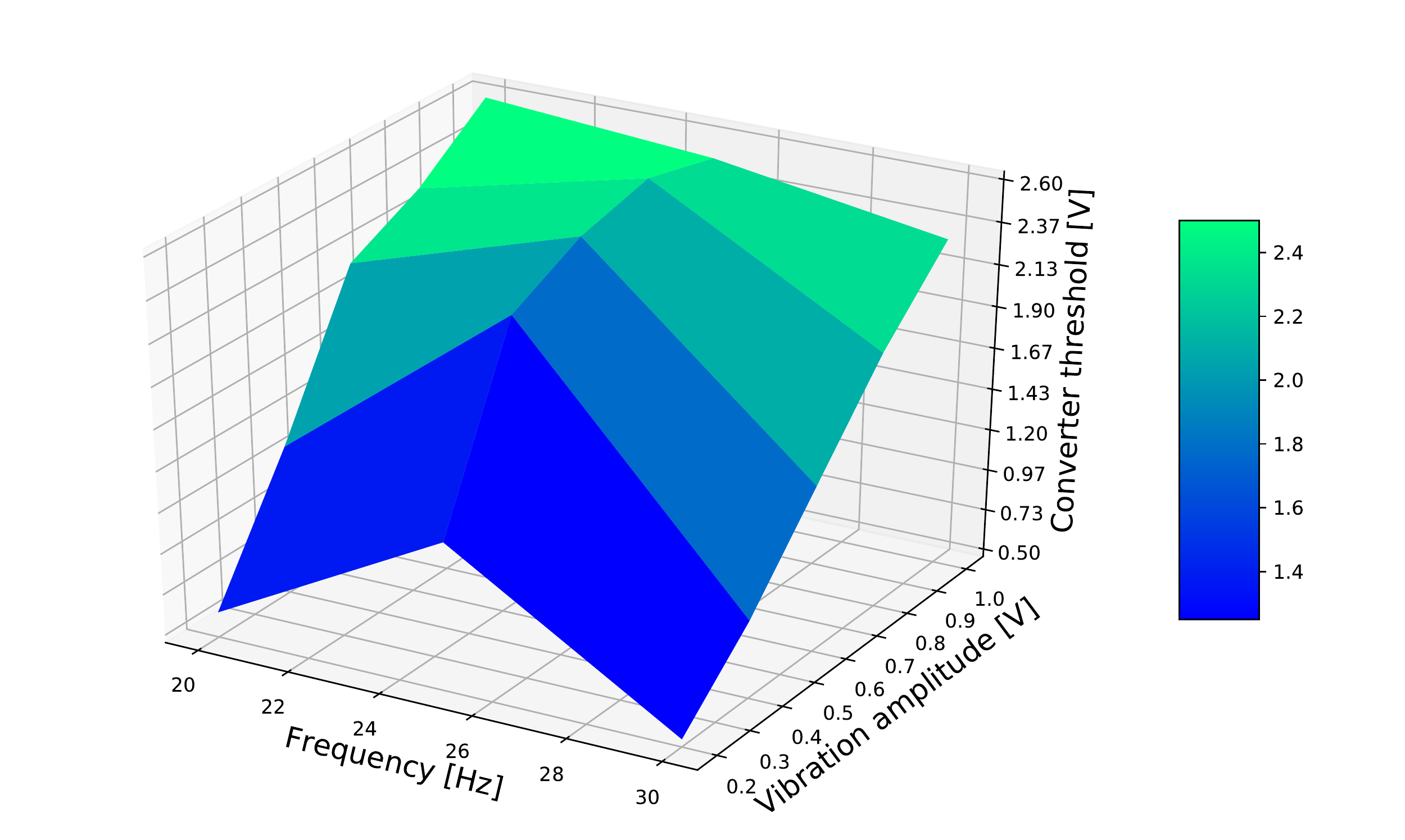}
\caption{The threshold voltage of the converter at which maximum power is harvested is a function of the vibration characteristics}
\label{fig:threshold_oc_setting_3d}
\end{figure}

Recalling, the intermittent execution pattern from Fig.~\ref{fig:transiently_powered_sensors}, we note that the capacitor voltage is always in the range of turn-on and turn-off thresholds under normal operating conditions.
From Eq.~\ref{eq:envel_formation}, we know that the rectified transducer operating voltage equals the capacitor voltage.
Thus, without a DC-DC converter, the transducer is operated at voltages between \SI{2.18}{\volt} and \SI{3.38}{\volt} in our prototype.
However, Fig.~\ref{fig:threshold_oc_setting_3d} shows that the \ac{mpp} voltage is often significantly lower.
With the converter-less design, the transducer is operated far from the \ac{mpp}, which can significantly reduce the output power as shown in figure \ref{fig:piezo_iV_curve_mppt}.
In contrast, by using a DC-DC converter, we can optimize the operating point of the transducer and as a result achieve greatly improved energy yield even with static \ac{mpp} tracking.

\subsection{Harvested power stored in the capacitor}
\label{Comparison_CL_CB}
We employ converter-less and converter-based designs to harvest power at varying levels of input vibrations. Fig.~\ref{fig:CL_CB_comparing_shaker} shows the harvested power calculated from the stored energy in the capacitor at varying frequency and amplitude levels of input vibrations. For the converter-based design, the harvested power is at the \ac{mpp}. Fig.~\ref{fig:CL_CB_comparing_shaker} shows that both designs harvest highest power at the resonance frequency due to the higher output power from the transducer. However, the converter-based circuit harvests higher power than the converter-less design at all frequency and amplitude levels of input vibrations. It offers about 26.5 to 409 times (on average, 112 times) higher power than the converter-less design while operating at the resonance frequency under varying vibration amplitudes. The reason for higher harvested power in converter-based design is the operation of the piezoelectric transducer at its \ac{mpp}.
\begin{figure}[t!]
\centering
\includegraphics[width=9cm, height=3.5cm]{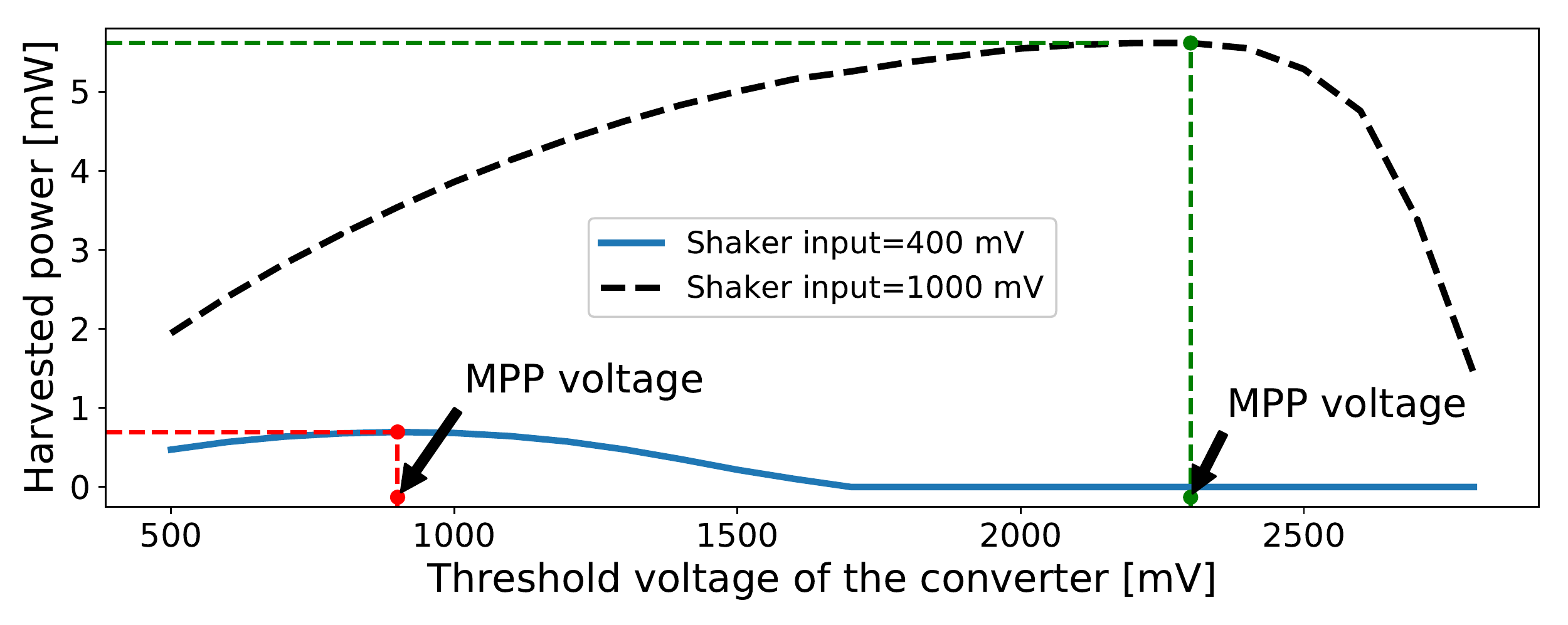}
\caption{Adjusting the threshold voltage of converter according to \ac{mpp} of the transducer allows to harvest higher energy}
\label{fig:efficiency}
\vspace{-0.5cm}
\end{figure}

\subsection{Impact of threshold voltage of DC-DC converter on the harvested power}
\label{threshold_DC_DC_converter}
We calculate the harvested power stored in the capacitor in the converter-based design using different voltage thresholds of the DC-DC converter and plot the results in Fig.~\ref{fig:efficiency}. It shows that at a specific voltage threshold of the DC-DC converter, higher power is harvested which corresponds to the \ac{mpp} of the transducer. Fig.~\ref{fig:efficiency} also shows that increasing the vibration amplitude (from \SI{400}{\milli\volt} to \SI{1000}{\milli\volt}) generates a higher transducer voltage, which in turn shifts the dynamic \ac{mpp} towards a higher voltage as discussed in Section~\ref{subsec:mpp_transducer}. The \ac{mpp} voltage depends on the underlying vibration characteristics. Fig.~\ref{fig:threshold_oc_setting_3d} plots the voltage threshold of DC-DC converter at \ac{mpp} with varying vibration frequency and amplitude levels. It shows that the threshold voltage of the converter at which maximum power is attained (i.e., \ac{mpp}), increases with the escalation in the vibration amplitude. For a given vibration level, the \ac{mpp} threshold is highest at the resonance frequency of the piezoelectric transducer due to the higher transducer voltage at this specific frequency.

\subsection{Power consumption of DC-DC converter}
\label{energy_consumption_DC_DC_converter}

In the converter-less design, the rectifier is the only significant source of losses.
The actual charging of the capacitor is almost lossless.
In the converter-based design instead, the converter itself causes significant losses, affecting the overall efficiency of the system.
Even if the converter is not operational, for example, when there is no vibration exciting the transducer, it still consumes a so-called quiescent current.
The ultra low power DC-DC boost converter\footnote{\label{note2}http://www.ti.com/lit/ds/symlink/bq25504.pdf} TI BQ25504 that we use in our prototype, consumes a maximum of \SI{4.2}{\micro\watt} of power at a typical capacitor voltage of \SI{3}{\volt}.
This is in the same range as the sleeping current of state of the art microcontrollers.
While the converter is actively charging the capacitor, it consumes additional power that, together with various switching losses reduces the efficiency, i.e. the ratio of output power to input power
The datasheet shows that the efficiency of the converter generally increases with the increase in input voltage and input current from the transducer.
For example, at typical harvesting conditions of \SI{1}{\volt} and \SI{100}{\micro\ampere}, the converter reaches at an efficiency of more than \SI{80}{\percent}.

Despite these losses, we find that the gain from operating the transducer at the \ac{mpp} by far outweighs the losses incurred by the DC-DC converter.

\section{Conclusion and Future Work}
\label{Conclusion_and_Future_Work}
\ac{keh} converts the ambient vibration, motion and mechanical energy into electrical energy to power the \ac{iot} sensor nodes. The conventional energy harvesting circuit employs a rectifier which rectifies the harvesting voltage to charge a capacitor which is further used to drive a load/node. However, this configuration lacks the operation of the transducer at its \ac{mpp} which results in lower harvested energy.
In this paper, we explore the \ac{mpp} of the \ac{keh} transducer for \textit{optimal} energy harvesting. We fix the voltage from the transducer to its \ac{mpp} resulting in higher harvested energy. 
We experimentally analyze the dynamic \ac{iv} characteristics and based on that specify the relation between the \ac{mpp} sampling rate and harvesting efficiency, outlining the need for real, dynamic \ac{mpp} tracking which is not currently available on commercial off-the-shelf hardware. We study the performance of converter-less and converter-based energy harvesting circuits in a controlled environment using a stable load shaker. The results show that the converter-based circuit harvests, on average, at least one order of magnitude higher power, than the converter-less design, due to its operation at \ac{mpp}, under a wide set of conditions typically found in human-centric applications. 

In future, we intend to study the performance of converter-less and converter-based circuits in a real world scenario using real-time \ac{mpp} tracking. In addition, an energy efficient circuit can be designed that solely operates on the harvested energy resulting in \ac{eno} of the batteryless \ac{iot} sensor node.

\bibliographystyle{IEEEtran}

\begin{thebibliography}{10}
\providecommand{\url}[1]{#1}
\csname url@samestyle\endcsname
\providecommand{\newblock}{\relax}
\providecommand{\bibinfo}[2]{#2}
\providecommand{\BIBentrySTDinterwordspacing}{\spaceskip=0pt\relax}
\providecommand{\BIBentryALTinterwordstretchfactor}{4}
\providecommand{\BIBentryALTinterwordspacing}{\spaceskip=\fontdimen2\font plus
\BIBentryALTinterwordstretchfactor\fontdimen3\font minus
  \fontdimen4\font\relax}
\providecommand{\BIBforeignlanguage}[2]{{%
\expandafter\ifx\csname l@#1\endcsname\relax
\typeout{** WARNING: IEEEtran.bst: No hyphenation pattern has been}%
\typeout{** loaded for the language `#1'. Using the pattern for}%
\typeout{** the default language instead.}%
\else
\language=\csname l@#1\endcsname
\fi
#2}}
\providecommand{\BIBdecl}{\relax}
\BIBdecl

\bibitem{lin2017survey}
J.~Lin{,} \emph{et~al.}, ``A survey on internet of things: Architecture,
  enabling technologies, security and privacy, and applications,'' \emph{IEEE
  Internet Things J.}, vol.~4, no.~5, pp. 1125--1142, 2017.

\bibitem{hester2017future}
J.~Hester{,} \emph{et~al.}, ``The future of sensing is batteryless,
  intermittent, and awesome,'' in \emph{Proc. 15th ACM SenSys}, 2017, p.~21.

\bibitem{estrada2018multiple}
J.~J. Estrada-L{\'o}pez{,} \emph{et~al.}, ``Multiple input energy harvesting
  systems for autonomous iot end-nodes,'' \emph{J. Low Power Electron. Appl.},
  vol.~8, no.~1, p.~6, 2018.

\bibitem{kuang2017energy}
Y.~Kuang{,} \emph{et~al.}, ``Energy harvesting during human walking to power a
  wireless sensor node,'' \emph{Sens. Actuator A-Phys.}, vol. 254, pp. 69--77,
  2017.

\bibitem{zhao2014shoe}
J.~Zhao{,} \emph{et~al.}, ``A shoe-embedded piezoelectric energy harvester for
  wearable sensors,'' \emph{Sensors}, vol.~14, no.~7, pp. 12\,497--12\,510,
  2014.

\bibitem{magno2016kinetic}
M.~Magno{,} \emph{et~al.}, ``Kinetic energy harvesting: Toward autonomous
  wearable sensing for internet of things,'' in \emph{Proc. 2016 IEEE SPEEDAM},
  2016, pp. 248--254.

\bibitem{ryokai2014energybugs}
K.~Ryokai{,} \emph{et~al.}, ``Energybugs: energy harvesting wearables for
  children,'' in \emph{Proc. 2014 ACM CHI}, 2014, pp. 1039--1048.

\bibitem{olivo2010kinetic}
J.~Olivo{,} \emph{et~al.}, ``A kinetic energy harvester with fast start-up for
  wearable body-monitoring sensors,'' in \emph{Proc. 4th
  PervasiveHealth}.\hskip 1em plus 0.5em minus 0.4em\relax IEEE, 2010, pp.
  1--7.

\bibitem{xie2014human}
L.~Xie{,} \emph{et~al.}, ``Human motion: sustainable power for wearable
  electronics,'' \emph{IEEE Pervasive Comput.}, vol.~13, no.~4, pp. 42--49,
  2014.

\bibitem{xiang2013powering}
T.~Xiang{,} \emph{et~al.}, ``Powering indoor sensing with airflows: a trinity
  of energy harvesting, synchronous duty-cycling, and sensing,'' in \emph{Proc.
  11th ACM SenSys}, 2013, p.~16.

\bibitem{khalifa2017harke}
S.~Khalifa{,} \emph{et~al.}, ``Harke: Human activity recognition from kinetic
  energy harvesting data in wearable devices,'' \emph{IEEE Trans. Mobile
  Comput.}, vol.~17, no.~6, pp. 1353--1368, 2017.

\bibitem{hemminki2013accelerometer}
S.~Hemminki{,} \emph{et~al.}, ``Accelerometer-based transportation mode
  detection on smartphones,'' in \emph{Proc. 11th ACM SenSys}, 2013, p.~13.

\end{thebibliography}

\end{document}